\documentclass[3p]{elsarticle}
\usepackage{lineno,hyperref}
\usepackage{graphicx}
\usepackage{amsmath}
\usepackage{color}
\usepackage[normalem]{ulem}

\journal{arXiv}
\begin{document}

\begin{frontmatter}

\title{Combined Mutiplicative-Heston Model for Stochastic Volatility}

\author[mymainaddress]{M. Dashti Moghaddam}
\author[mymainaddress]{R. A. Serota\fnref{myfootnote}}
\fntext[myfootnote]{serota@ucmail.uc.edu}

\address[mymainaddress]{Department of Physics, University of Cincinnati, Cincinnati, Ohio 45221-0011}

\begin{abstract}
We consider a model of stochastic volatility which combines features of the multiplicative model for large volatilities and of the Heston model for small volatilities. The steady-state distribution in this model is a Beta Prime and is characterized by the power-law behavior at both large and small volatilities. We discuss the reasoning behind using this model as well as consequences for our recent analyses of distributions of stock returns and realized volatility.
\end{abstract}

\begin{keyword}
 Volatility \sep Heston \sep Multiplicative \sep Beta Prime \sep Distribution Tails
\end{keyword}

\end{frontmatter}

\section{Introduction\label{Introduction}}
Distributions of stock returns (SR) have long fascinated researchers -- see \cite{harris2001empirical} for a good summary of earlier works, dating back to Mandelbrot in the early 60s. It is widely believed that, at least for daily returns, SR distributions have fat (power-law) tails. Accordingly, SR distributions were fitted with a number of fat-tailed distributions, such as stable, Student's $t$, and generalized $t$ -- see \cite{harris2001empirical} and references therein and a more recent \cite{rathie2012stable}. Studies of intra-day returns also argue in favor of the power-law-tail hypothesis \cite{gerig2009model,behfar2016long}. Alternatively, the long multi-day returns seem to be described just as well with an exponentially decaying distribution \cite{dragulescu2002probability,liu2017distributions}.

Student's $t$ distribution has an appeal of being underpinned by a simple multiplicative stochastic volatility model \cite{nelson1990arch}, which leads to an Inverse Gamma (IGa) steady-state distribution for the variance of the volatility\cite{praetz1972distribution,fuentes2009universal,ma2014model,liu2017distributions}. Its drawback, however, is that IGa decays exponentially quickly for small values of volatility. Another widely used stochastic volatility model is the Heston model \cite{heston1993closed,dragulescu2002probability}, which leads to a Gamma (Ga) steady-state distribution for the variance \cite{dragulescu2002probability,liu2017distributions}. Ga scales as power law for small volatilities and serves as an underpinning for the exponentially decaying SR distribution that may also be suitable for fitting multi-day returns. In fact, the Kolmogorov-Smirnov (KS) test does not give a clear advantage to either multiplicative or Heston model \cite{liu2017distributions}.

In this paper we propose a stochastic volatility model that marries the properties of multiplicative and Heston models and results in a Beta Prime (BP) steady-state distribution that replicates the power-law properties of Ga for small volatilities and of IGa for large volatilities. We discuss what consequences this model has for our previous results on SR distributions \cite{liu2017distributions} and realized variance \cite{dashti2018implied}. We also consider the question of whether the stochastic equation for SR should be understood in Stratonovich or Ito context \cite{perello2000black}.

This paper is organized as follows. In Section \ref{ModelsV} we introduce the combined multiplicative-Heston model and discuss its steady-state BP distribution. In Section \ref{SR} we discuss Startonovich versus Ito interpretation of the SR equation and its consequences for SR and leverage. We discuss SR distribution fitting and its moments in light of the new model. In Section \ref{RV} we discuss the theoretical value of the variance of realized variance (RV) \cite{dashti2018implied} for this model and compare it with the numerical results from the market data.

\section{Models of Volatility\label{ModelsV}}

The two widely used mean-reverting models of stochastic volatility $\sigma_t$, expressed in terms of stochastic variance $v_t=\sigma_t^2$, are multiplicative (MM) \cite{ma2014model}
\begin{equation}
\mathrm{d}v_t = -\gamma(v_t - \theta)\mathrm{d}t + \kappa_M v_t\mathrm{d}W_t^{(2)}
\label{IGa}
\end{equation}
and Heston (HM) \cite{dragulescu2002probability}
\begin{equation}
\mathrm{d}v_t = -\gamma(v_t - \theta)\mathrm{d}t + \kappa_H \sqrt{v_t}\mathrm{d}W_t^{(2)}
\label{Ga}
\end{equation}
where $\mathrm{d}W_t^{(2)}$ is the normally distributed Wiener process, $\mathrm{d}W_t^{(2)} \sim \mathrm{N(}0,\, \mathrm{d}t \mathrm{)}$. The steady-state distributions for $v_t$ and $\sigma_t$ are respectively $\mathrm{IGa(}v_t;\, \frac{\alpha }{\theta}+1,\, \alpha \mathrm{)}$ and $2\sigma_t \cdot \mathrm{IGa(}\sigma_t^2;\, \frac{\alpha }{\theta}+1,\, \alpha \mathrm{)}$ for MM and $\alpha \mathrm{Ga(}\alpha v_t;\, \alpha,\, \theta \mathrm{)}$ and $2\sigma_t \cdot \alpha \mathrm{Ga(}\alpha \sigma_t^2;\, \alpha,\, \theta \mathrm{)}$ for HM, where 
\begin{equation}
\alpha = \frac{2\gamma \theta}{\kappa_{M,H}^2}
\label{alpha}
\end{equation}
for both models, with $\alpha > 1$ for the HM \cite{dragulescu2002probability,ma2014model,liu2017distributions}. A simple relationship exists between $\kappa_M$ and $\kappa_H$: $\kappa_M \theta \approx \kappa_H \sqrt{\theta}$ or $\kappa_H^2 / \kappa_M^2 \approx  \theta$.

Here we introduce a new combination multiplicative-Heston model (MHM), given by 
\begin{equation}
\mathrm{d}v_t = -\gamma(v_t - \theta)\mathrm{d}t + \sqrt{\kappa_M^2 v_t^2 + \kappa_H^2 v_t }\mathrm{d}W_t^{(2)}
\label{sdrGaGa}
\end{equation}
Its steady-state distribution is a BP, 
\begin{equation}
BP(v_t; p,q,\beta)=\frac{(1+\frac{v_t}{\beta})^{-p-q}(\frac{v_t}{\beta})^{-1+p}}{\beta B(p,q)}
\label{BP}
\end{equation}
where $B(p,q)$ is the beta function,
\begin{equation}
\label{p}
p=\frac{2 \gamma \theta}{\kappa_H^2}
\end{equation}
and
\begin{equation}
\label{q}
q=1+\frac{2 \gamma}{\kappa_M^2}
\end{equation}
are the shape parameters and 
\begin{equation}
\label{beta}
\beta=\frac{\kappa_H^2}{\kappa_M^2}
\end{equation}
is the scale parameter and, according to the above, $ \beta \approx \theta = \alpha(\theta/\alpha) $ is the product of the scale parameters of MM and HM. The limiting behaviors of BP is 
\begin{equation}
BP(v_t; p,q,\beta) \propto (\frac{v_t}{\beta})^{-q-1}, \hspace{.25cm} v_t \gg \beta
\label{BPlargev}
\end{equation}
and
\begin{equation}
BP(v_t; p,q,\beta) \propto (\frac{v_t}{\beta})^{p-1}, \hspace{.25cm} v_t \ll \beta
\label{BPsmallv}
\end{equation}
that is the same as in MM and HM respectively ($p > 1$ for the latter). Furthermore, for $p \gg 1$, BP can mimic IGa for $v_t \ll \beta$ and, for $q \gg 1$, BP can mimic Ga for $v_t \gg \beta$. Stochastic volatility is, accordingly, distributed as $2 \sigma_t BP(\sigma_t^2; p,q,\beta)$.

We would like to point out that, obviously, BP has an extra shape parameter relative to IGa and Ga and that it is non-trivial to extract the latter two from BP as limits. We also point out that ordinarily it is assumed that the equations for stochastic variance should be understood in the Ito sense. However, since the term that couples to the Gaussian noise contains powers of $v_t$ it is appropriate to ponder a Stratonovich interpretation as well. We observe, however, that for MHM transition from Stratonovich to Ito involves a simple renormalization of constants $\gamma$ and $\theta$ -- or just one of them in its MM and HM limits.

\section{Stock Returns Distributions and Moments\label{SR}}

The standard equation for the stock price reads \cite{osborne1964random}
\begin{equation}
\frac{\mathrm{d}S_t}{S_t} = \mu \mathrm{d}t + \sigma_t \mathrm{d}W_t^{(1)}
\label{dSStr}
\end{equation}
where $\mathrm{d}W_t^{(1)} \sim \mathrm{N(}0,\, \mathrm{d}t \mathrm{)}$. As is for volatility, this equation is almost always interpreted as Ito but a Stratonovich interpretation also needs explored. In the latter case, the transformation to Ito yields
\begin{equation}
\frac{\mathrm{d}S_t}{S_t} = \mu \mathrm{d}t + \frac{\sigma_t^2}{2}\mathrm{d}t+ \sigma_t \mathrm{d}W_t^{(1)}
\label{dSIto}
\end{equation}
Using Ito calculus, eqs. (\ref{dSStr}) and (\ref{dSIto}) can be rewritten as 
\begin{equation}
\mathrm{d}\log{S_t} = \mu \mathrm{d}t  - \frac{\sigma_t^2}{2}\mathrm{d}t + \sigma_t \mathrm{d}W_t^{(1)}
\label{dlogSStr}
\end{equation}
and
\begin{equation}
\mathrm{d}\log{S_t} = \mu \mathrm{d}t  + \sigma_t \mathrm{d}W_t^{(1)}
\label{dlogSIto}
\end{equation}
respectively.

The first conclusion that can be drawn is that the equation used to estimate the implied volatility index VIX \cite{demeterfi1999guide,demeterfi1999more}
\begin{equation}
\frac{\mathrm{d}S_t}{S_t} - \mathrm{d}\log{S_t} =\frac{\sigma_t^2}{2}\mathrm{d}t
\label{VIX}
\end{equation}
is not affected by which interpretation -- Ito or Stratonovich - is used. It is also obvious that the Black-Scholes equation is not affected either, since it assumes a constant -- or at least a non-stochastic -- volatility; see \cite{perello2000black} for a detailed analysis.

Denoting $r_t = \ln (S_t/S_0)$ and $x_t = r_t - \mu t$, the equations (\ref{dlogSStr}) and (\ref{dlogSIto}) for log returns become
\begin{equation}
\mathrm{d}x_t = -\frac{\sigma_t^2}{2}\mathrm{d}t + \sigma_t\mathrm{d}W_t^{(1)}
\label{xtStr}
\end{equation}
and 
\begin{equation}
\mathrm{d}x_t = \sigma_t\mathrm{d}W_t^{(1)} 
\label{xtIto}
\end{equation}
respectively. 

It should be pointed out that in general $\mathrm{d}W_t^{(1)}$ and $\mathrm{d}W_t^{(2)}$ are correlated as
\begin{equation}
\mathrm{d}W_t^{(2)} = \rho \mathrm{d}W_t^{(1)} + \sqrt{1-\rho^2} \mathrm{d}Z_t
\label{corr}
\end{equation}
Where $\mathrm{d}Z_t$ is independent of $\mathrm{d}W_t^{(1)}$, and $\rho \in [-1,\, 1]$ is the correlation coefficient. The latter can be evaluated from leverage correlations \cite{bouchaud2001leverage,perello2002stochastic}. We showed, howewver, that SR distributions are not effected by these correlations and that one can set $\rho=0$ \cite{liu2017distributions}. 

The SR distribution in (\ref{xtIto}) can be evaluated as the product distribution (PD) of volatility and normal distribution \cite{liu2017distributions}. We also showed that in (\ref{xtStr}) the first term in the r.h.s. does not yield significant corrections to the SR distribution until very long periods of returns  \cite{liu2017distributions}. Nonetheless, we evaluated the distribution in (\ref{xtStr}) as a joint probability (JP) distribution and found that PD fit of the market data had lower KS values than JP \cite{liu2017distributions}, which points to that (\ref{dSStr}) should be interpreted as Startonovich and reduced to Ito as according to (\ref{dSIto}). Furthermore, comparing (\ref{dSStr}) and (\ref{dlogSIto}), if the former is interpreted as Stratonovich, it is the latter that should be used for evaluations of leverage. In \cite{dashti2018correlations}, we show that indeed this approach gives a better statistical fit of the market data.

Using now $2 \sigma_t BP(\sigma_t^2; p,q,\beta)$ as the distribution of stochastic volatility $\sigma_t$ and taking a PD with the normal distribution in (\ref{xtIto}) in a manner explained in \cite{liu2017distributions}, we obtain the following SR distribution:
\begin{equation}
\psi_{MH}(z) = \frac{\Gamma \left(q+\frac{1}{2}\right) U\left(q+\frac{1}{2}, \frac{3}{2}-p, \frac{z^2}{2 \beta  \tau}\right)}{\sqrt{2\pi \beta \tau} B \left(p,q \right)} 
\label{pdfMHM}
\end{equation}
where $U$ is the confluent hypergeometric function, $\mathrm{d}x_t$ was replaced with $z$ and and $\mathrm{d}t$ was replaced with $\tau$ -- the number of days over which the returns are calculated.

\begin{figure}[!htbp]
\centering
\begin{tabular}{cc}
\includegraphics[width = 0.5 \textwidth]{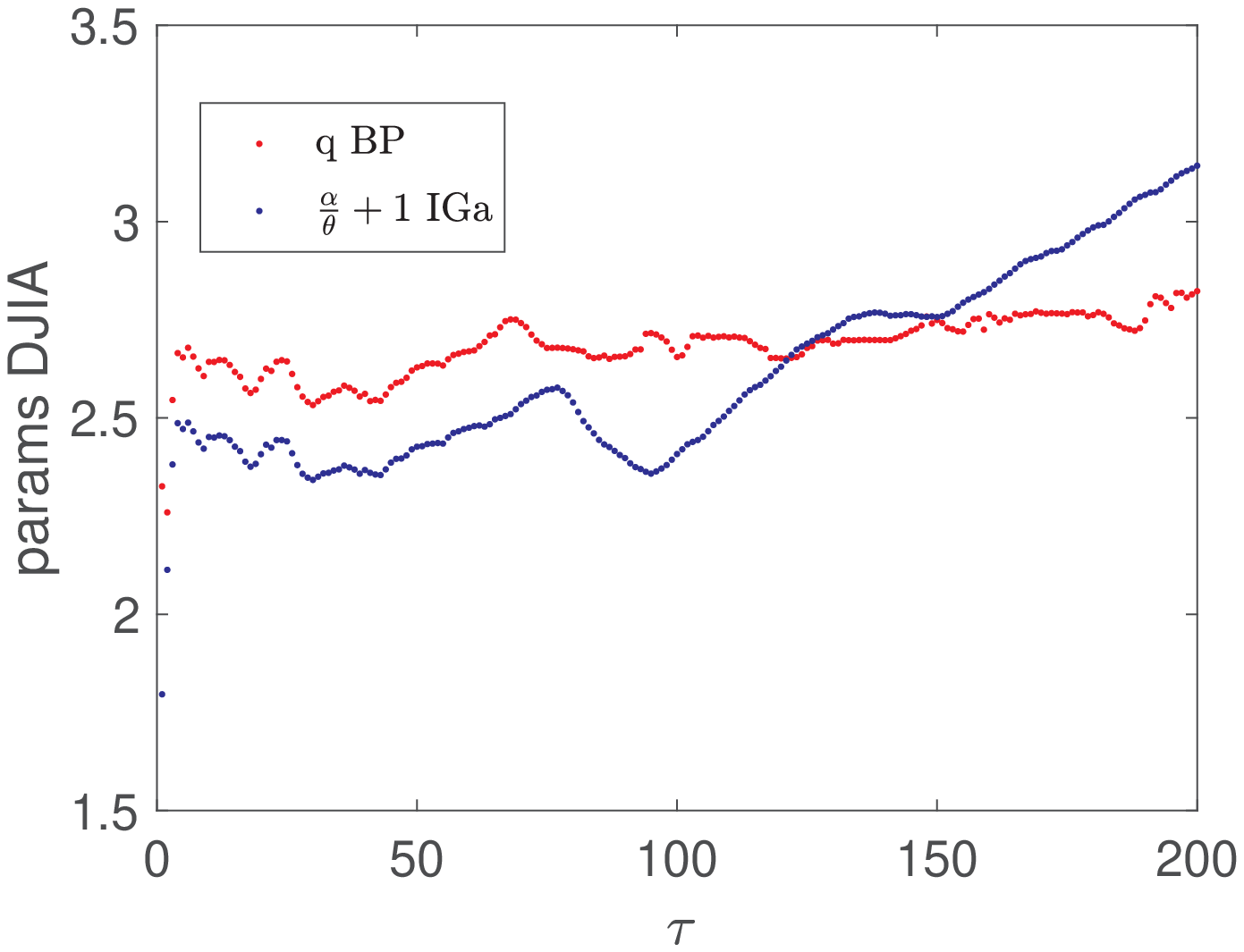}
\includegraphics[width = 0.5 \textwidth]{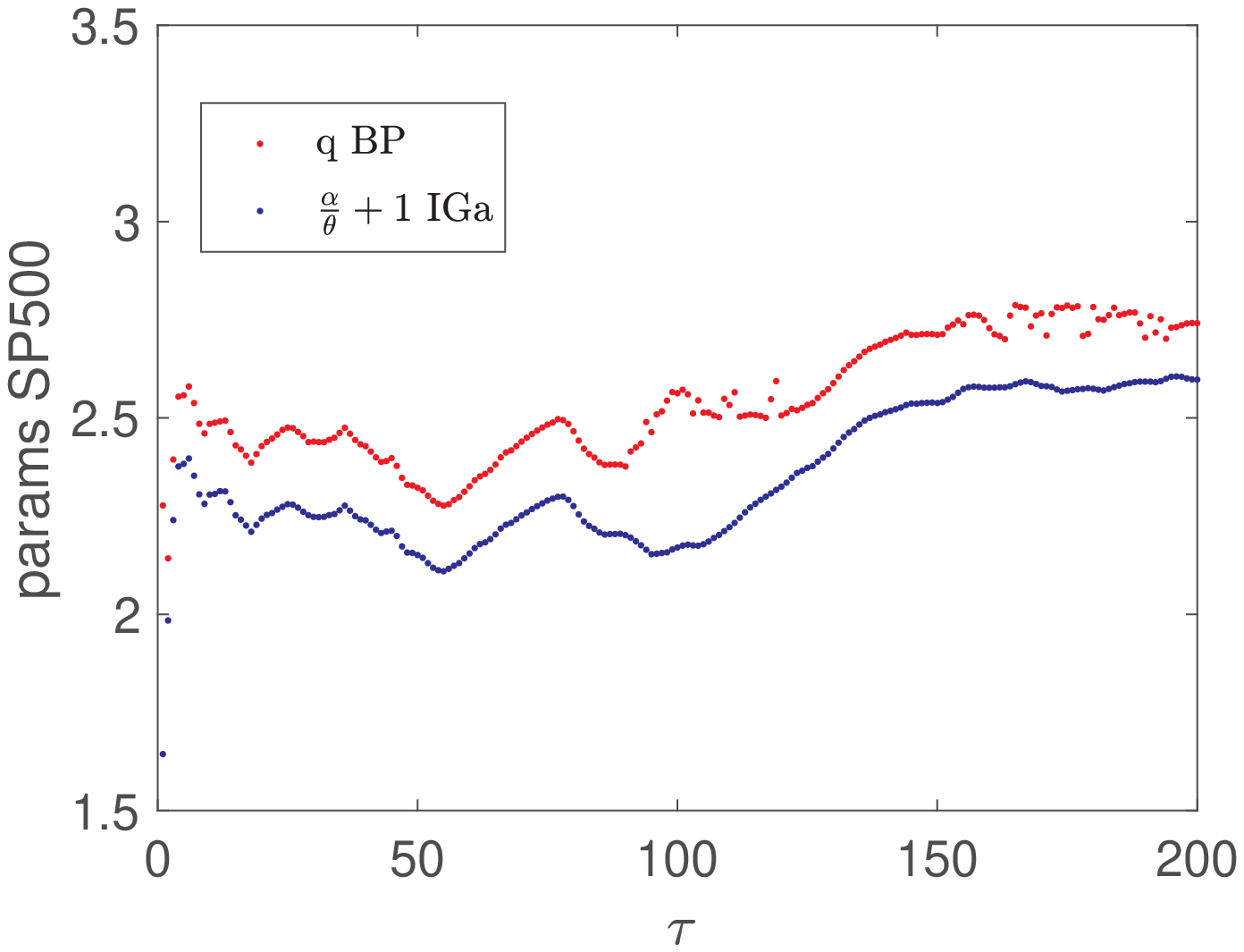}
\end{tabular}
\caption{Parameter $q$ in the BP distribution (\ref{BP}) versus parameter $\frac{\alpha}{\theta}+1$ in the IGa distribution}
\label{alphatheta-q}
\end{figure}

\begin{figure}[!htbp]
\centering
\begin{tabular}{cc}
\includegraphics[width = 0.5 \textwidth]{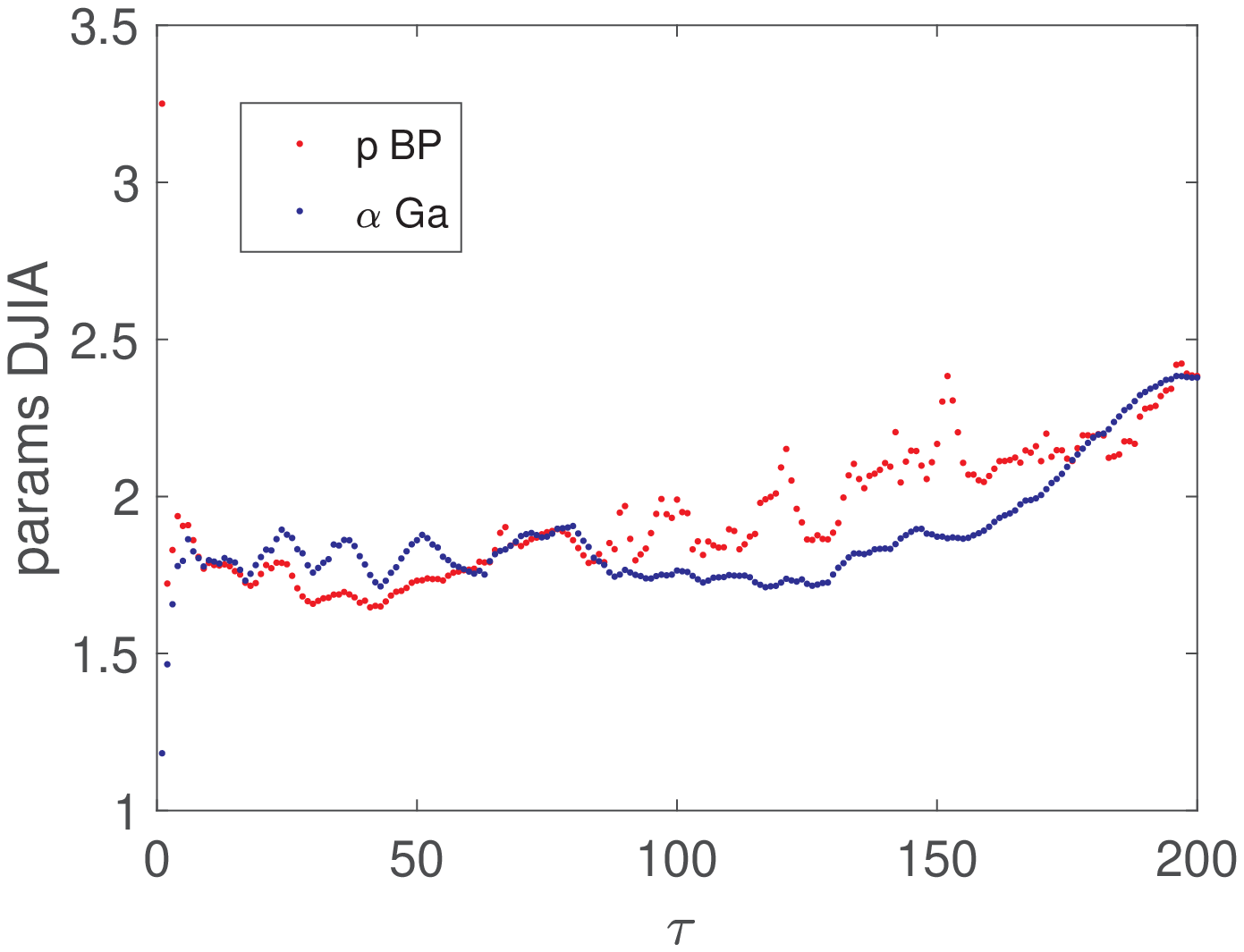}
\includegraphics[width = 0.5 \textwidth]{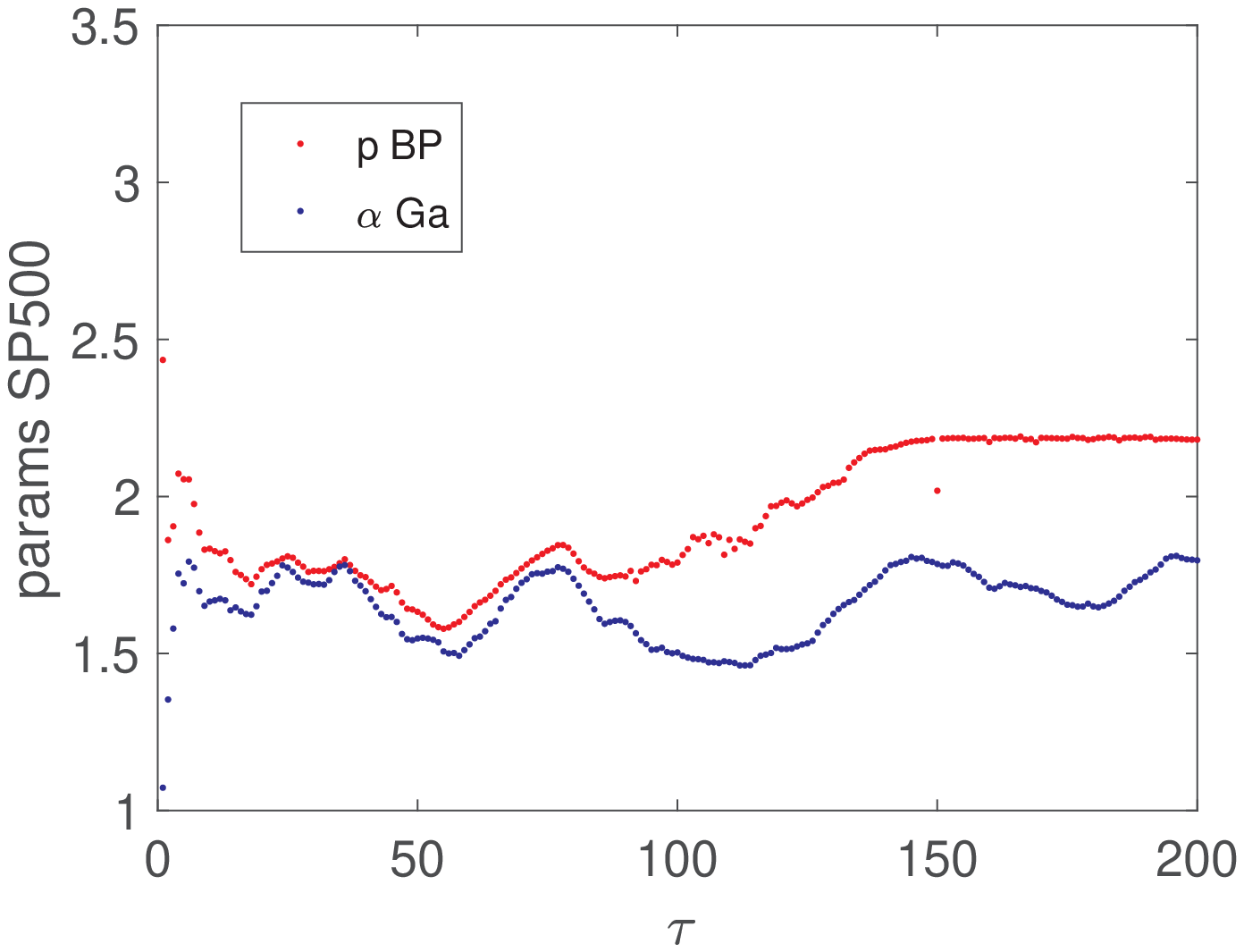}
\end{tabular}
\caption{Parameter $p$ in the BP distribution (\ref{BP}) versus parameter $\alpha$ in the Ga distribution}
\label{alpha-p}
\end{figure}

\begin{figure}[!htbp]
\centering
\begin{tabular}{cc}
\includegraphics[width = 0.5 \textwidth]{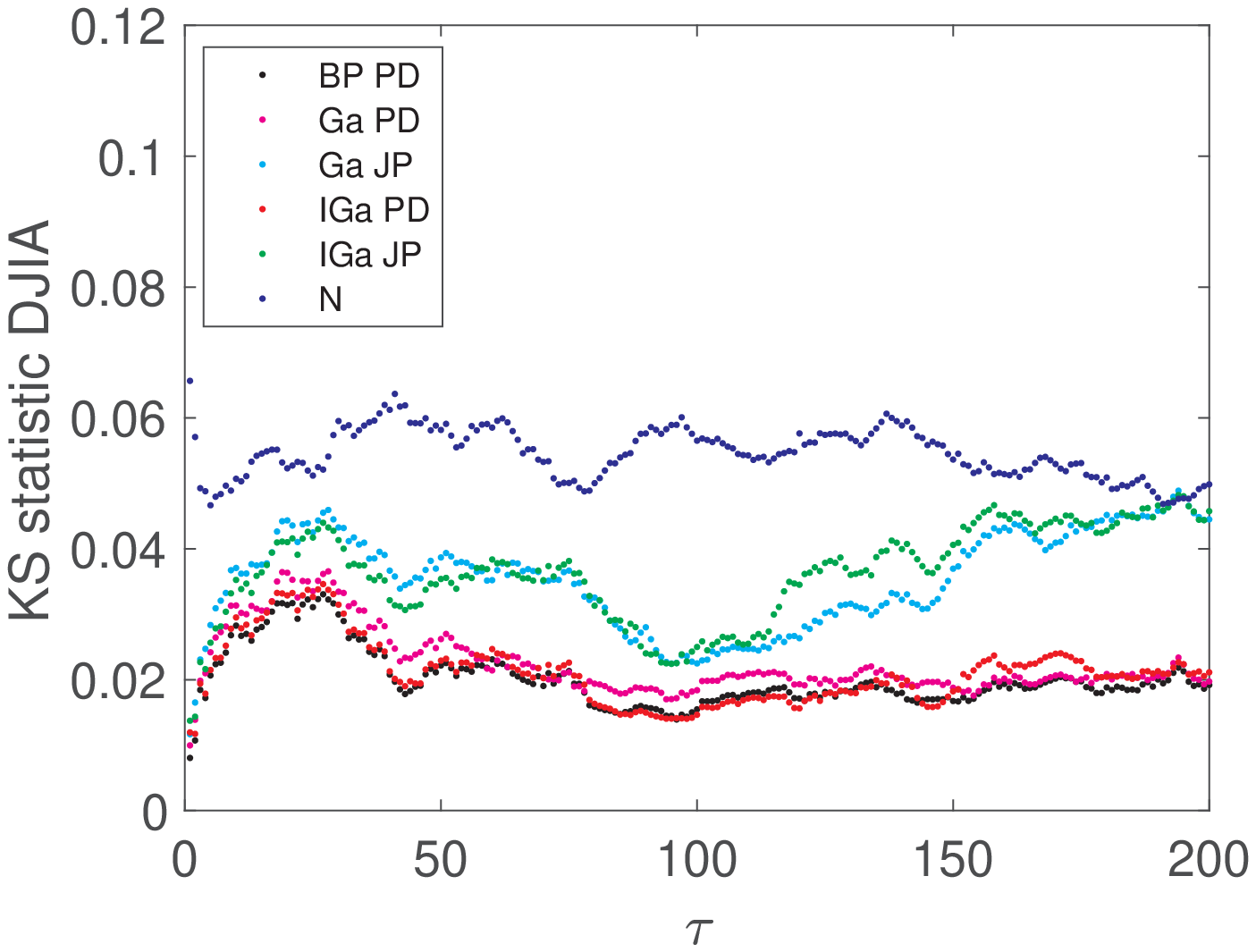}
\includegraphics[width = 0.5 \textwidth]{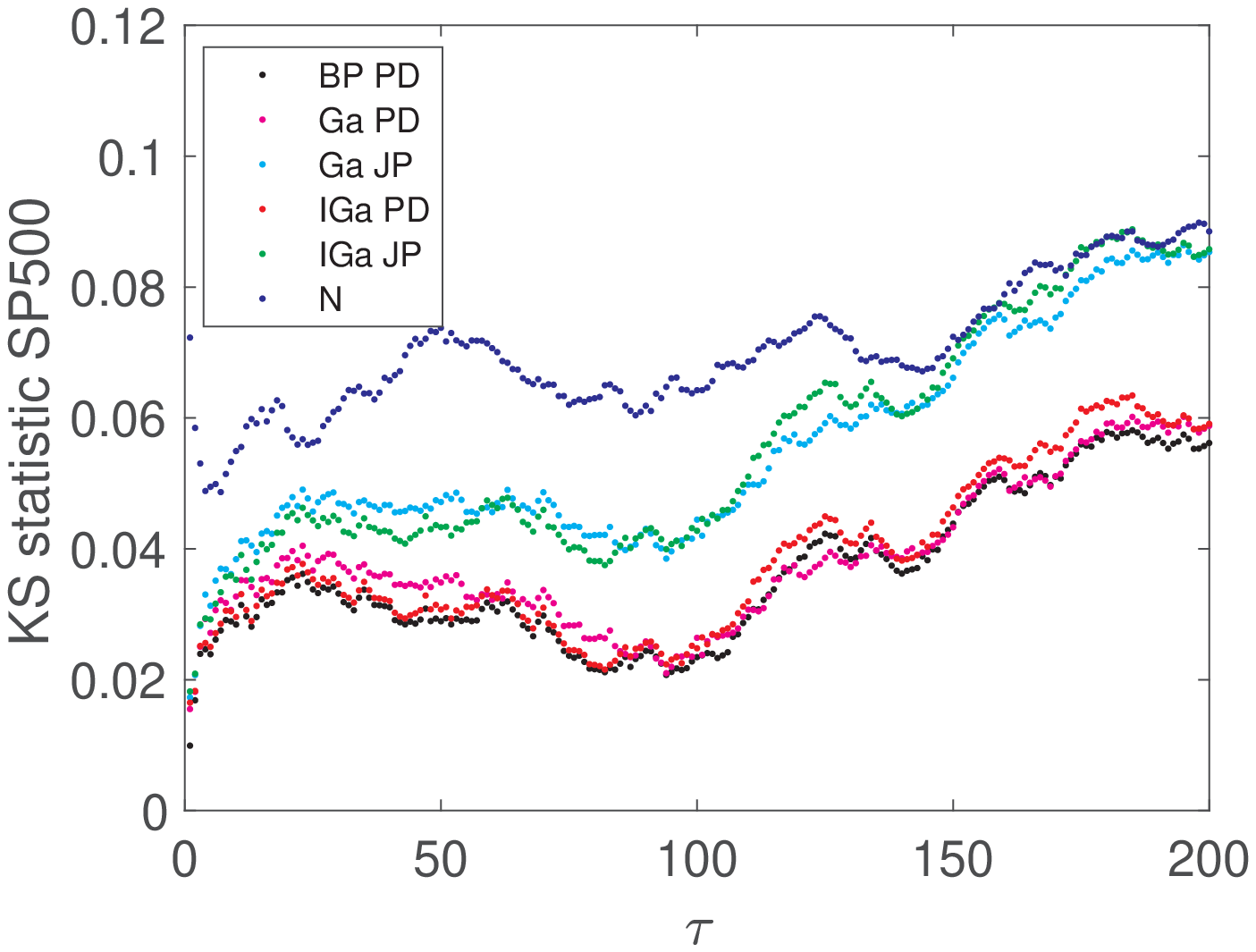}
\end{tabular}
\caption{KS test results. PD statistic is better than JP pointing to (\ref{dSIto}) as a preferred interpretation.}
\label{KS}
\end{figure}

\begin{figure}[!htbp]
\centering
\begin{tabular}{cc}
\includegraphics[width = 0.5 \textwidth]{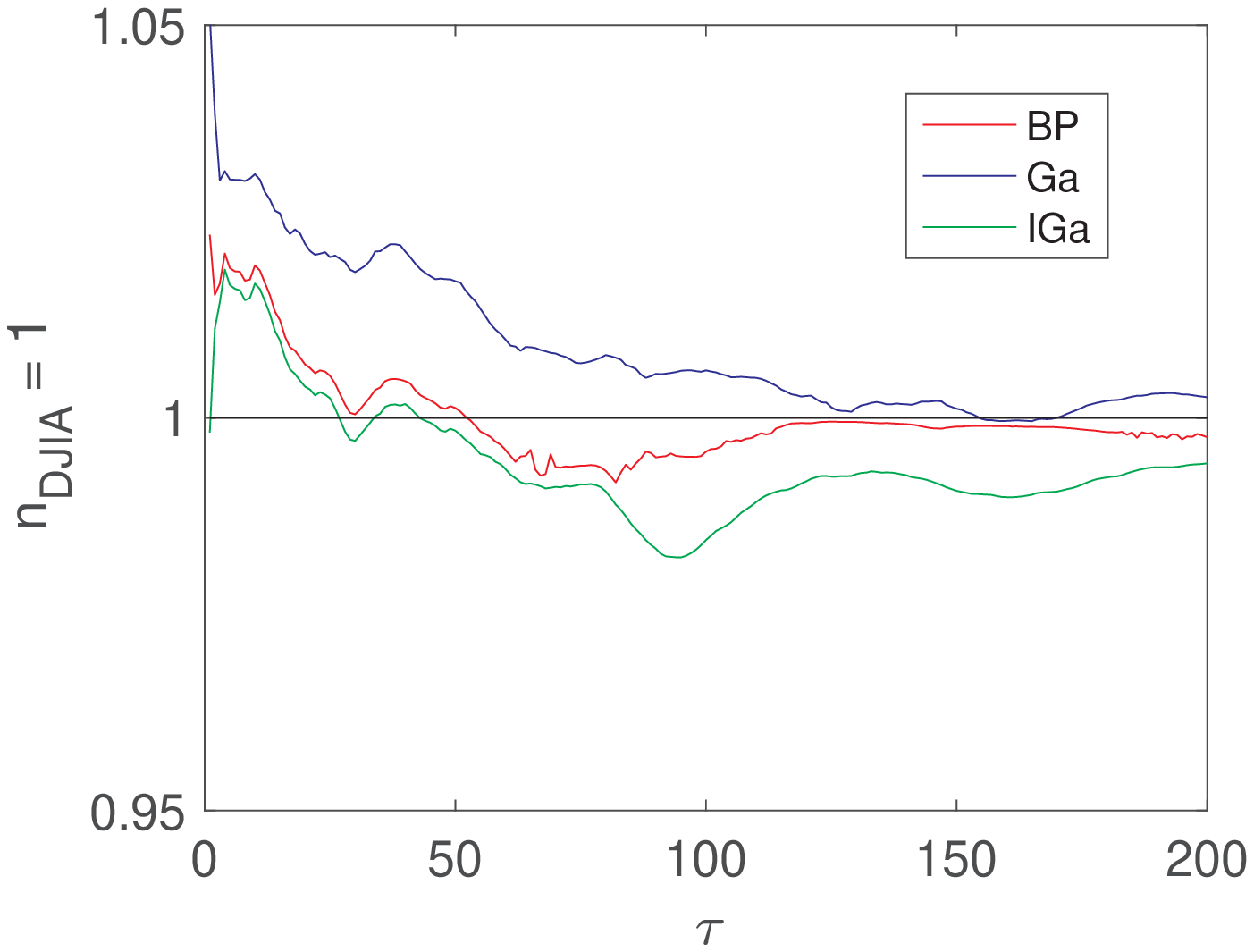}
\includegraphics[width = 0.5 \textwidth]{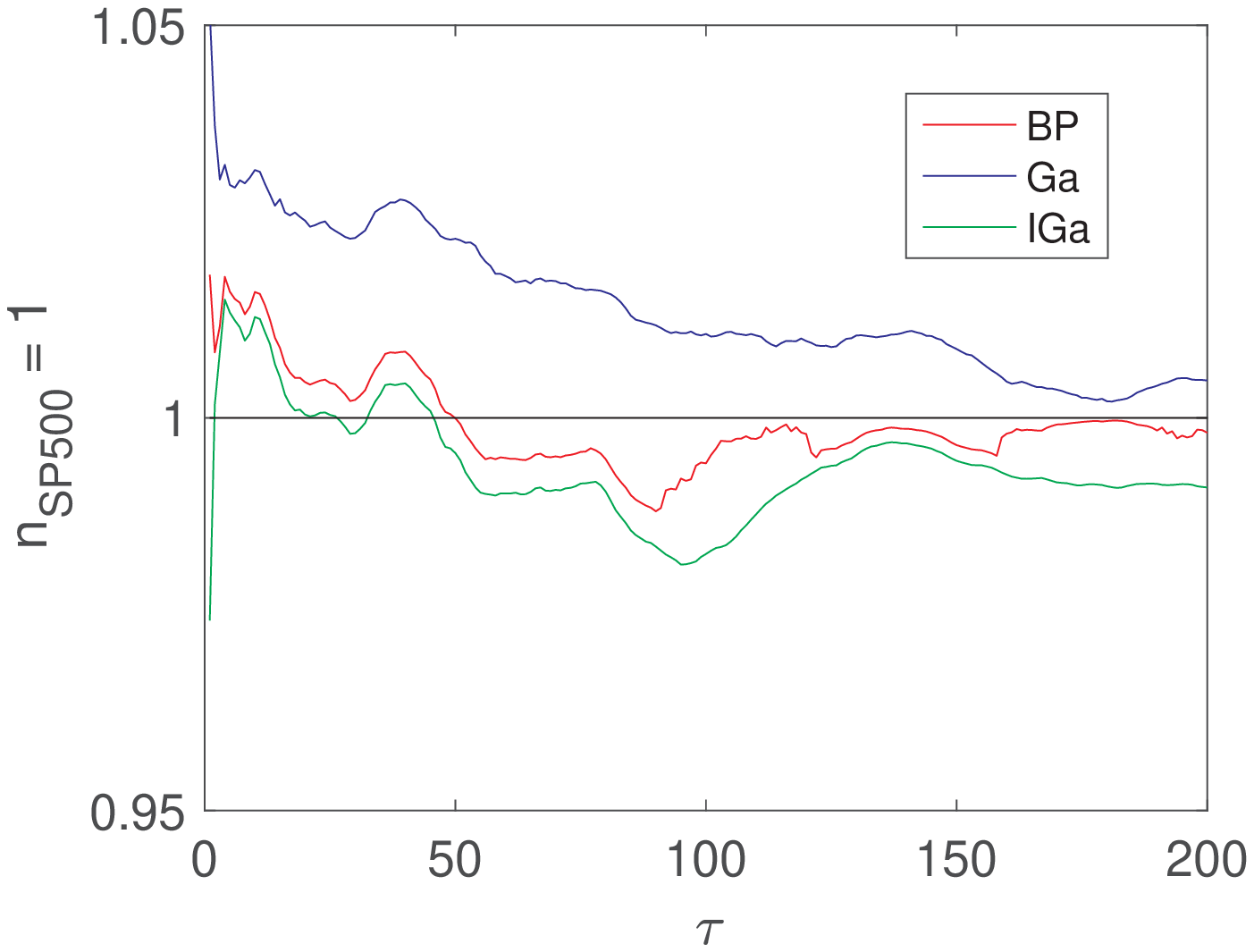}\\
\includegraphics[width = 0.5 \textwidth]{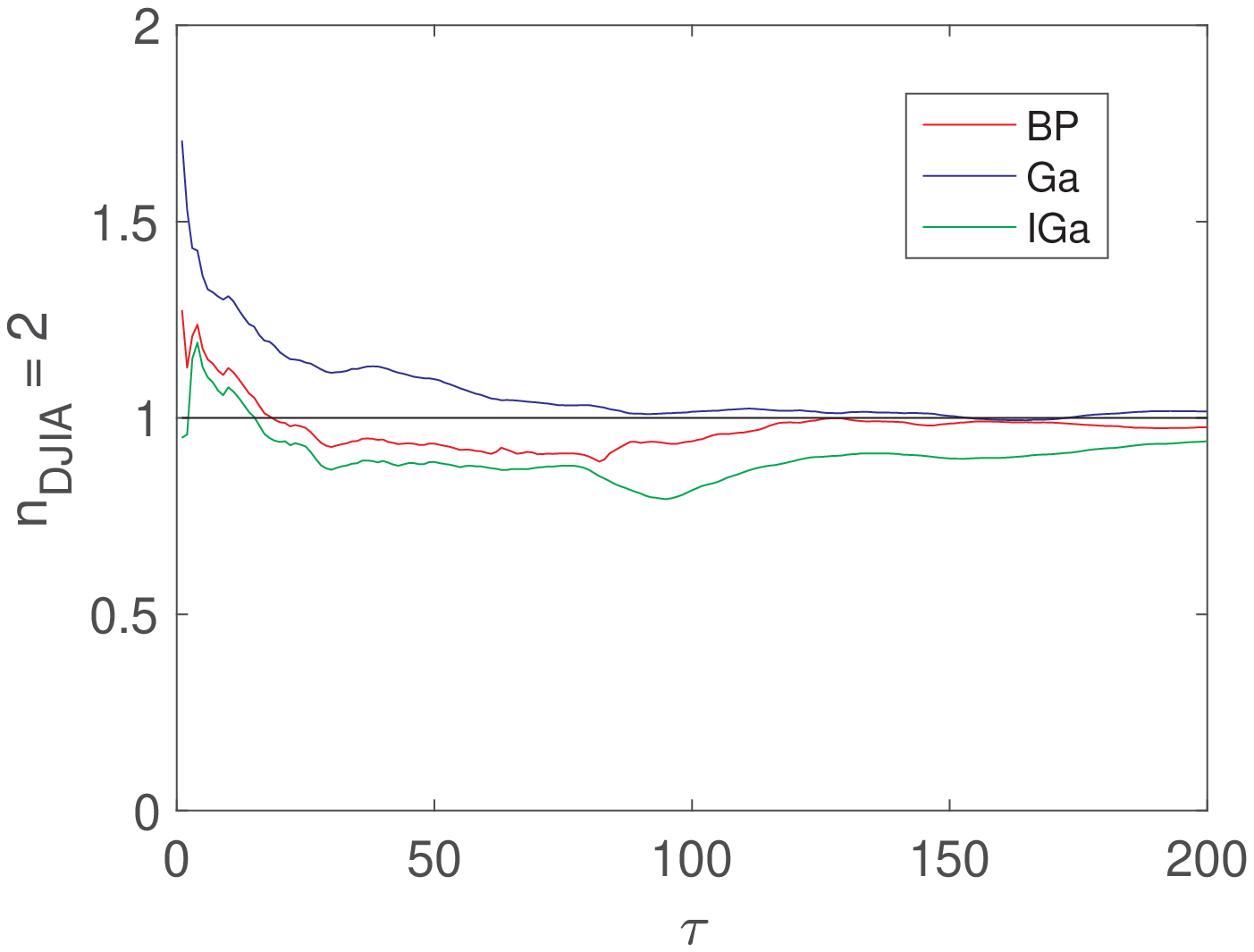}
\includegraphics[width = 0.5 \textwidth]{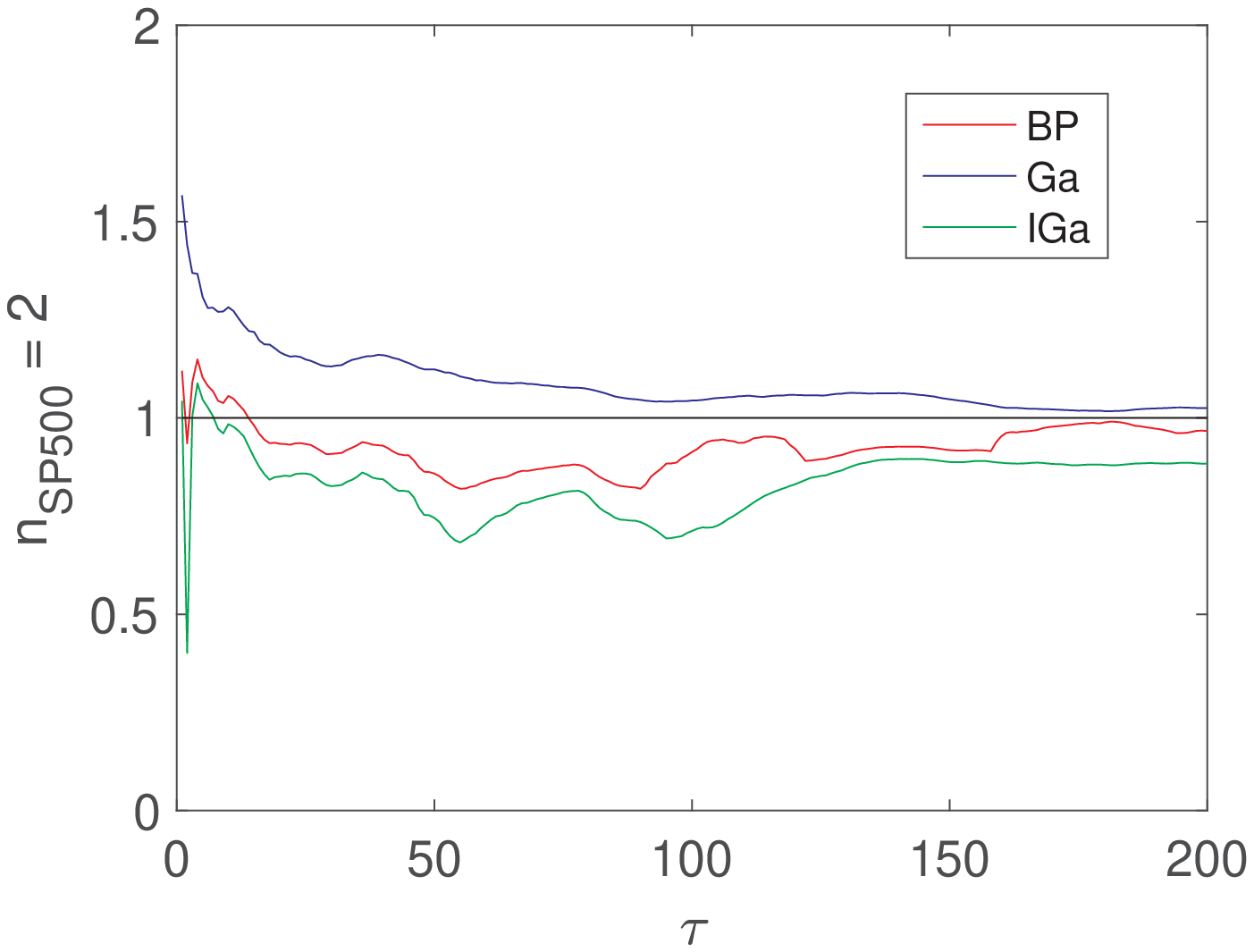}
\end{tabular}
\caption{Reduced moments of stock returns, per (\ref{moments}).}
\label{momentsfig}
\end{figure}

We use the same fitting procedures as in \cite{liu2017distributions} to find the parameters $p$, $q$ and $\beta$. Figs. \ref{alphatheta-q} - \ref{momentsfig} are plotted as a function of $\tau$, the number of days over which the returns are calculated. Fig. \ref{alphatheta-q} shows $q$ of MHM vis-a-vis $\frac{\alpha}{\theta}+1$ of MM, which reflects long tails of the stochastic variance per (\ref{BPlargev}). Fig. \ref{alpha-p} shows $p$ of MHM vis-a-vis $\alpha$ of HM, which reflects small stochastic variance behavior per (\ref{BPsmallv}). Fig. \ref{theta} shows $\theta$, the mean value of the stochastic variance, for all three models; for BP it is determined using (\ref{momentsMHM1}) below and for IGa and Ga from direct fitting \cite{liu2017distributions}. We also show $\theta$ calculated directly from the variance of SR, $\overline{z^{2}}=\theta \tau$ -- see (\ref{moments}) below -- and parameter $\beta$ to confirm (see above) that $ \beta \approx \theta$. Fig. \ref{KS} gives KS values for SR fits, the only new element relative to \cite{liu2017distributions} being the fit using (\ref{pdfMHM}).

Fig. \ref{momentsfig} contains reduced moments for $n=1$ and $n=2$
\begin{equation}
\left(\frac{\overline{z^{2n}}}{E(z^{2n})}\right)^{\frac{1}{2 n}}
\label{moments}
\end{equation}
where $\overline{z^{2n}}$ is numerically calculated average from the market data and $E(z^{2n})$ is its analytical value calculated from all three models. $E_M(z^{2n})$ and $E_H(z^{2n})$ are given in \cite{liu2017distributions} so here we only list the MHM values:
\begin{equation}
E_{MH}(z^{2})=\frac{p \beta \tau}{q-1} = \theta \tau
\label{momentsMHM1}
\end{equation}
\begin{equation}
E_{MH}(z^{4})=\frac{3 p (p+1)\beta^2 \tau^2}{(q-1)(q-2)}=\frac{3(2\gamma\theta^2+\kappa_H^2\theta) \tau^2}{2\gamma-\kappa_M^2} 
\label{momentsMHM2}
\end{equation}
We point out that one must have $2\gamma>\kappa_M^2$. We recall that in MM $\frac{2\gamma}{\kappa_M^2}=\frac{\alpha}{\theta}$ defines the exponent of the power-law tail. This parameter is greater than one \cite{liu2017distributions} -- see also (\ref{BPlargev}) and Fig. \ref{alphatheta-q}.

\begin{figure}[!htbp]
\centering
\begin{tabular}{cc}
\includegraphics[width = 0.5 \textwidth]{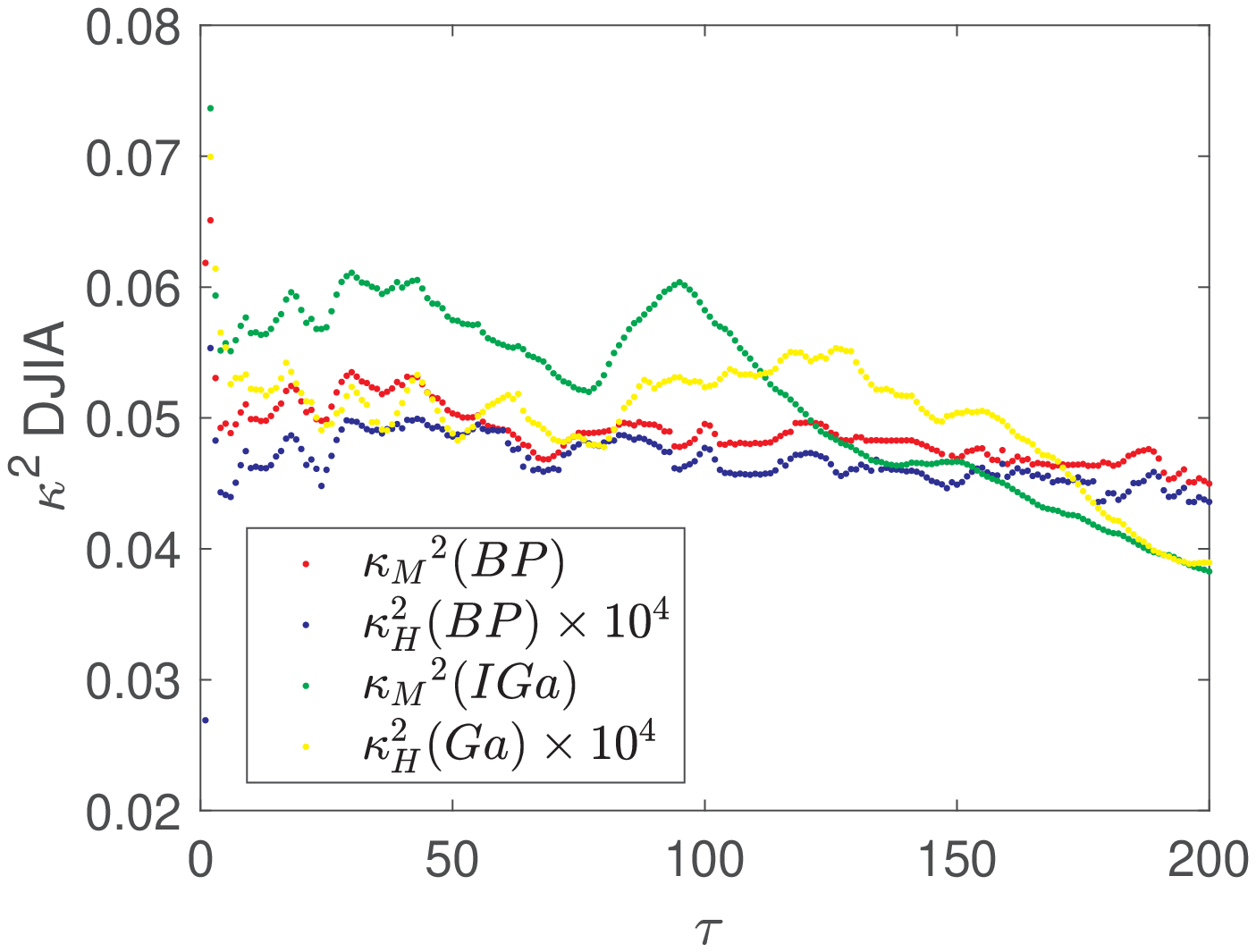}
\includegraphics[width = 0.5 \textwidth]{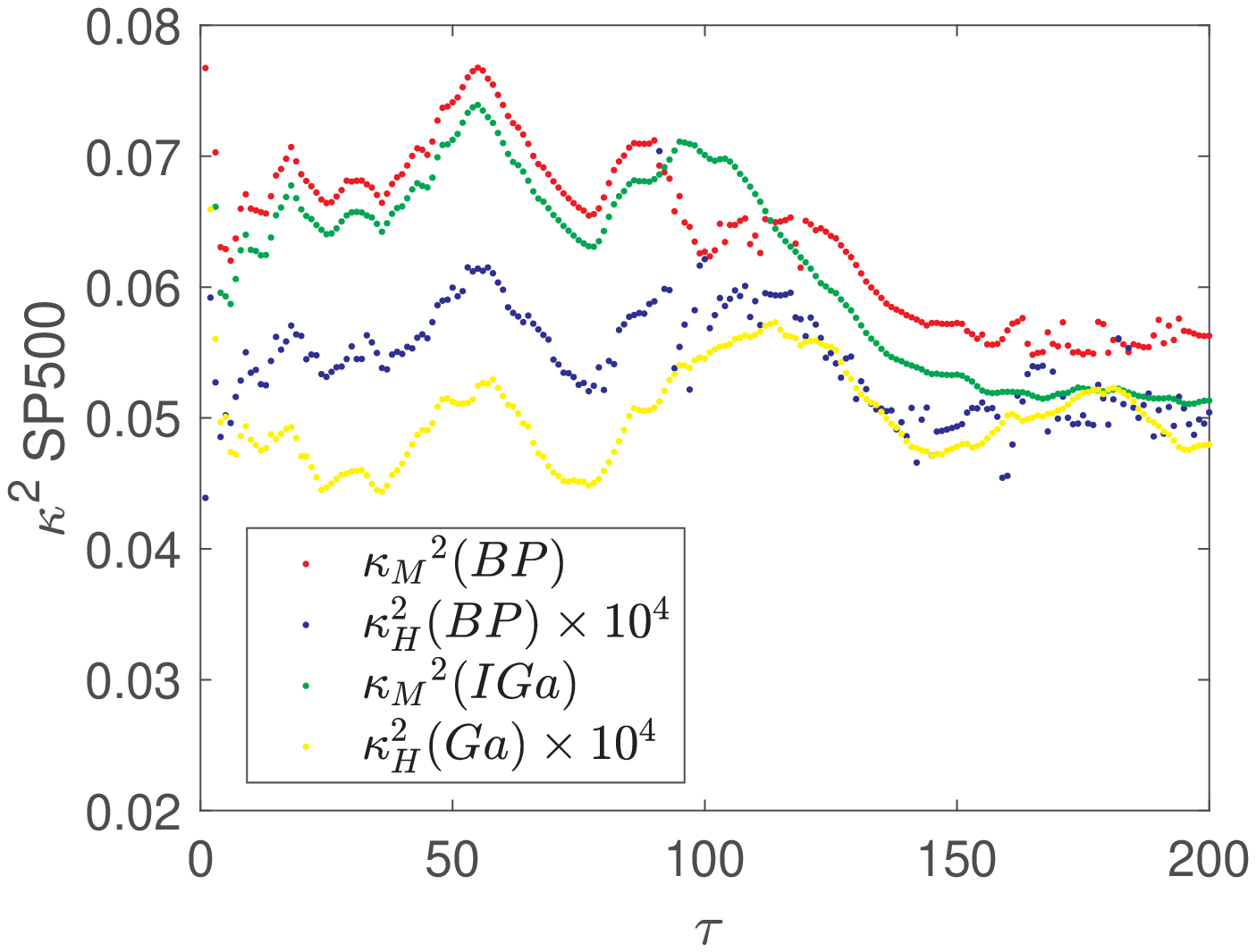}
\end{tabular}
\caption{$\kappa_M^2$ (BP) and $\kappa_H^2$ (BP) found from (\ref{q}) and (\ref{beta})} with $\gamma_{\text{DJIA}}=0.042$ and $\gamma_{\text{S\&P}}=0.041$ found by fitting market data correlation function -- see Sec. \ref{RV}. \newline $\kappa_M^2$ (IGa) and $\kappa_H^2$ (Ga) are values found using MM and HM respectively.
\label{kappas}
\end{figure}

JP results in Fig. \ref{KS} is shown to illustrate Stratonovich versus Ito discussion in this Section. While KS test does not give an advantage to either of the three models -- MM, HM, and MHM -- the latter describes the moments in Fig. \ref{momentsfig} clearly better than the other two. Finally, using (\ref{q}) and (\ref{beta}), we can find $\kappa_M$ and $\kappa_H$ once we know $\gamma$, which can be found by fitting the market data correlation function -- see Sec. \ref{RV}; $\kappa_M^2$ and $\kappa_H^2$ are plotted in Fig. \ref{kappas}.

\begin{figure}[!htbp]
\centering
\begin{tabular}{cc}
\includegraphics[width = 0.5 \textwidth]{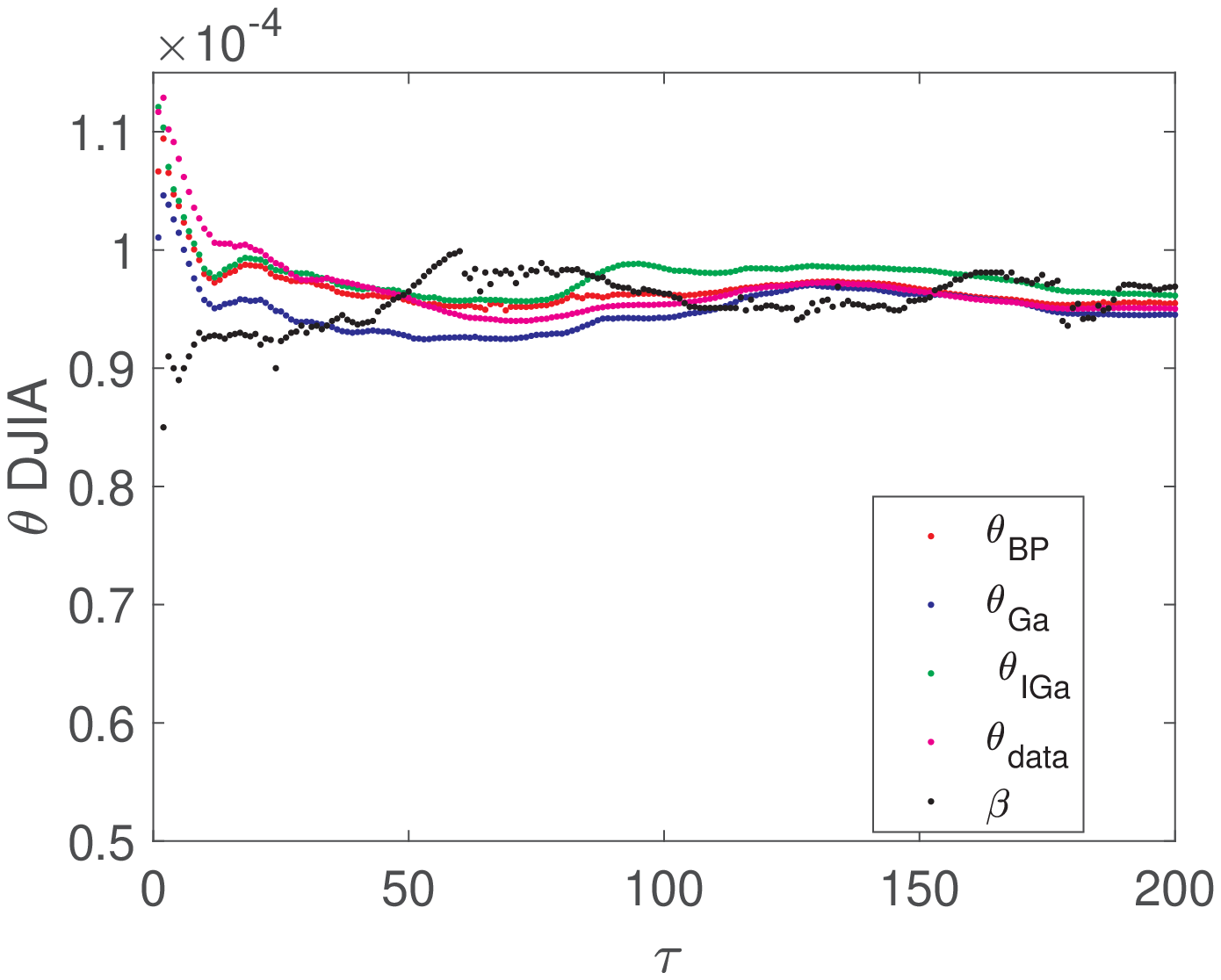}
\includegraphics[width = 0.5 \textwidth]{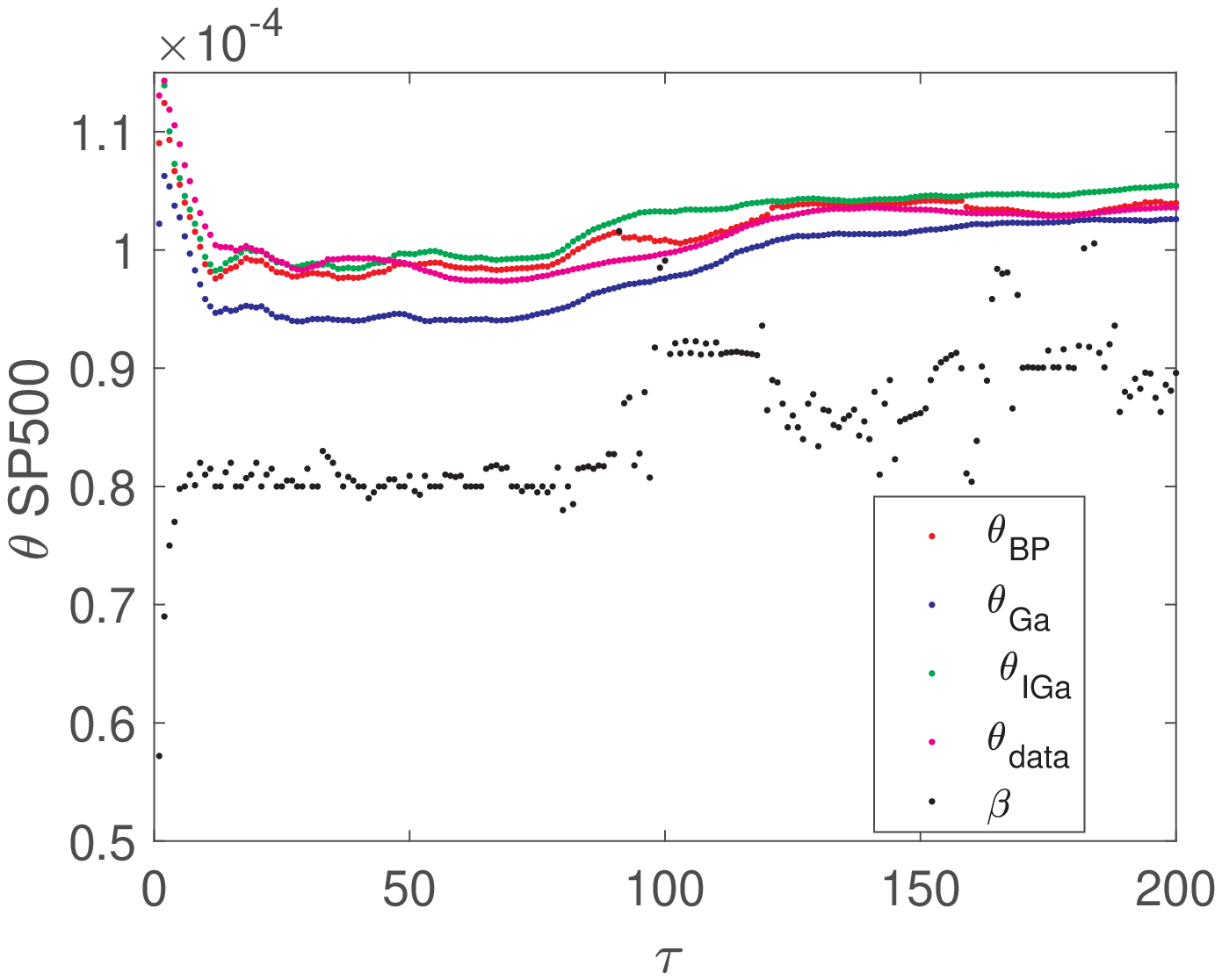}\\
\includegraphics[width = 0.5 \textwidth]{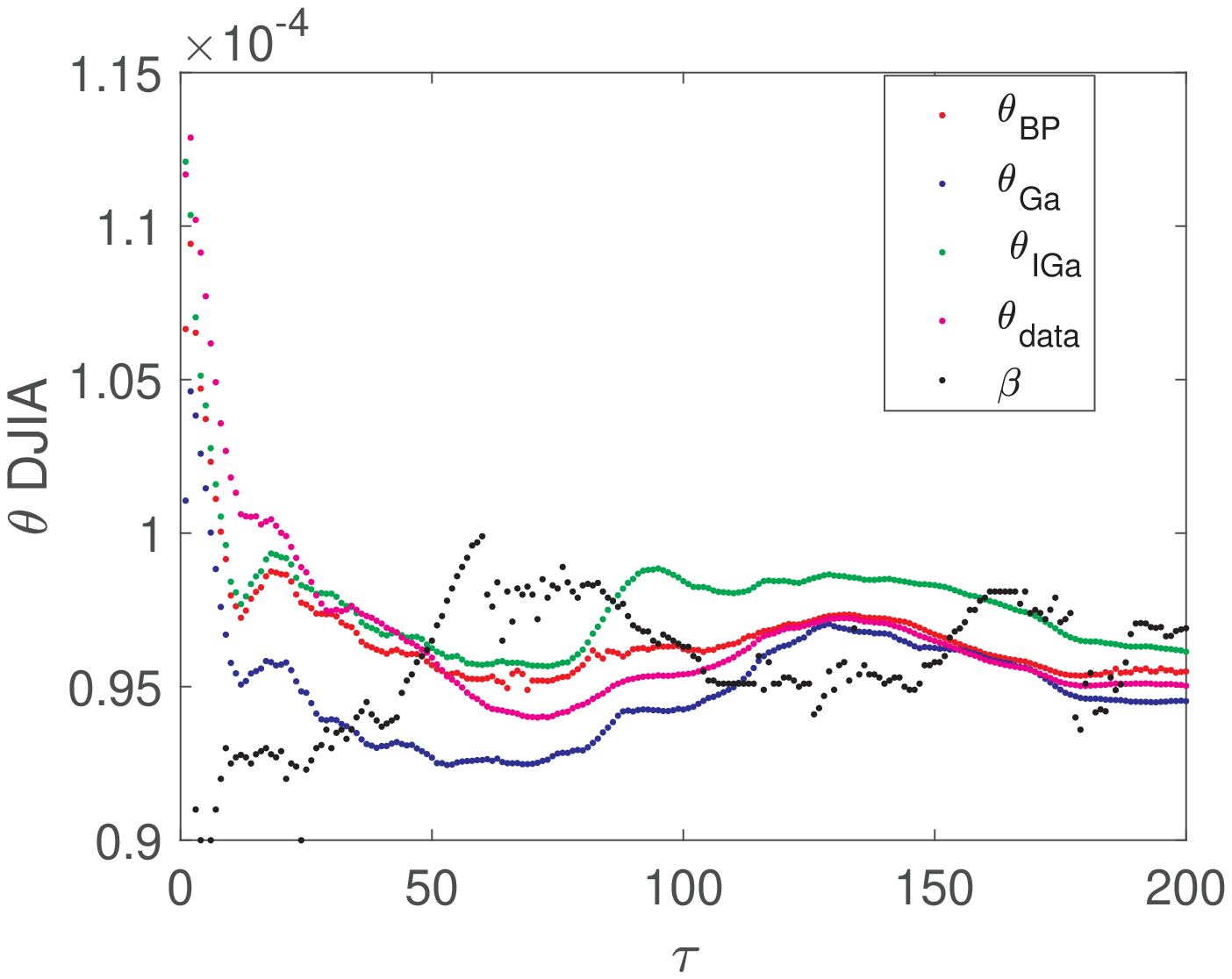}
\includegraphics[width = 0.5 \textwidth]{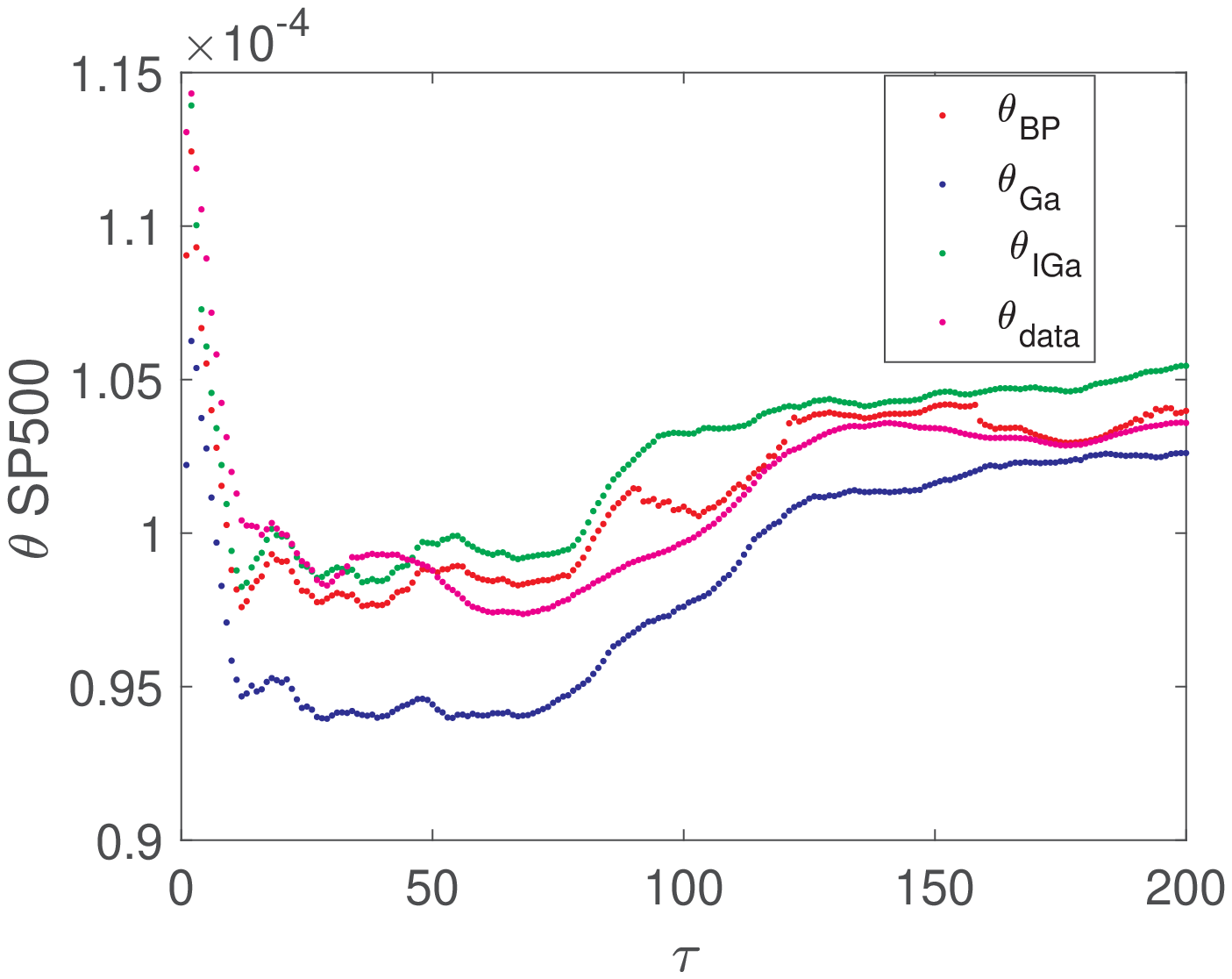}
\end{tabular}
\caption{Mean value of stochastic variance $\theta$ for MM, HM, MHM and data; $\beta \approx \theta$ also included.  Bottom row is the same as top row on smaller scale.}
\label{theta}
\end{figure}

\section{Application to Realized Variance \label{RV}} 
The correlation function of stochastic variance \cite{dashti2018correlations}
\begin{equation}
E[v_t v_{t+\tau}]=E[v_t]^2+var[v_t] e^{-\gamma \tau}
\label{meanVarCor}
\end{equation}
can be used (along with leverage \cite{bouchaud2001leverage,perello2002stochastic}, with minor differences in the result \cite{dashti2018correlations}), to determine $\gamma$. Here
\begin{equation}
E[v_t]=\theta 
\label{theta}
\end{equation}
for the mean-reverting models (for BP it can be obtained by integration with (\ref{BP}), and \begin{equation}
var[v_t]=E[v_t^2]-(E[v_t])^2
\label{meanVarVar}
\end{equation}
To find $E[v_t v_{t+\tau}]$ we must use $\mathrm{d}x_t^2 = v_t\mathrm{d}t$, which follows from (\ref{xtIto}). We observe that 
\begin{equation}
E[\mathrm{d}x_t^2 \mathrm{d}x_{t+\tau}^2 ]  = (E[v_t v_{t+\tau}] + 2 E[v_t ^2]) \mathrm{d}t^2
\label{dxt2corr}
\end{equation}
and in particular,
\begin{equation}
E[\mathrm{d}x_t^4] = 3 E[v_t ^2] \mathrm{d}t^2
\label{dxt4}
\end{equation}
The factor of 3 is purely combinatorial and is model-independent. It can be verified for any of the mentioned models. For instance, integrating $v_t^2$ with BP in (\ref{BP}), we find
\begin{equation}
E[v_t^2]=\frac{2\gamma\theta^2+\kappa_H^2\theta}{2\gamma-\kappa_M^2}
\label{meanVarSqu}
\end{equation}
in agreement with (\ref{momentsMHM2}) and (\ref{dxt4}). When higher moments exist, we can similarly obtain $E[\mathrm{d}x_t^{2n}] = (2n-1)!! E[v_t ^n] \mathrm{d}t^n$ -- see for instance those for HM in \cite{liu2017distributions}.

\begin{figure}[!htbp]
\centering
\begin{tabular}{cc}
\includegraphics[width = 0.5 \textwidth]{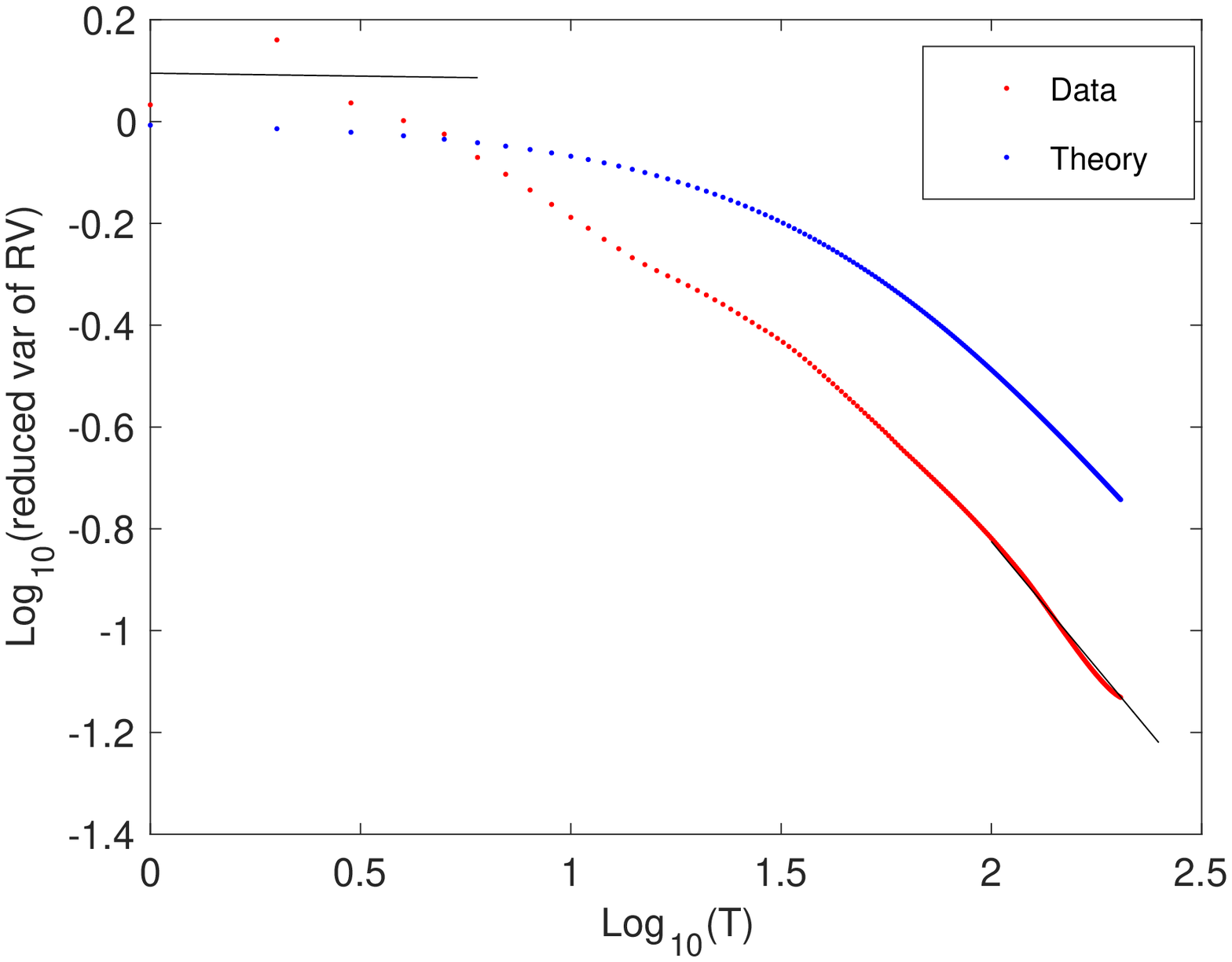}
\includegraphics[width = 0.5 \textwidth]{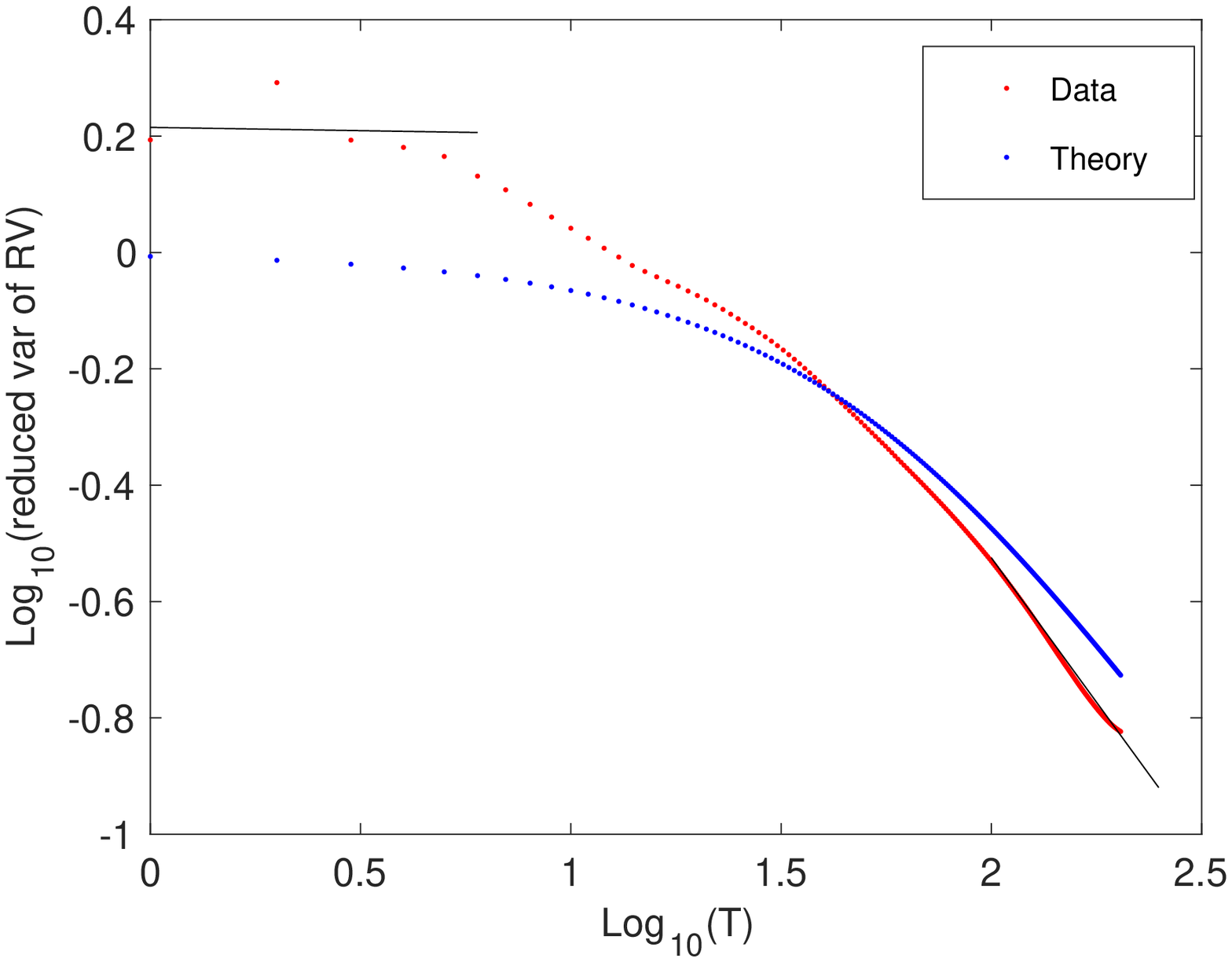}\\
\includegraphics[width = 0.5 \textwidth]{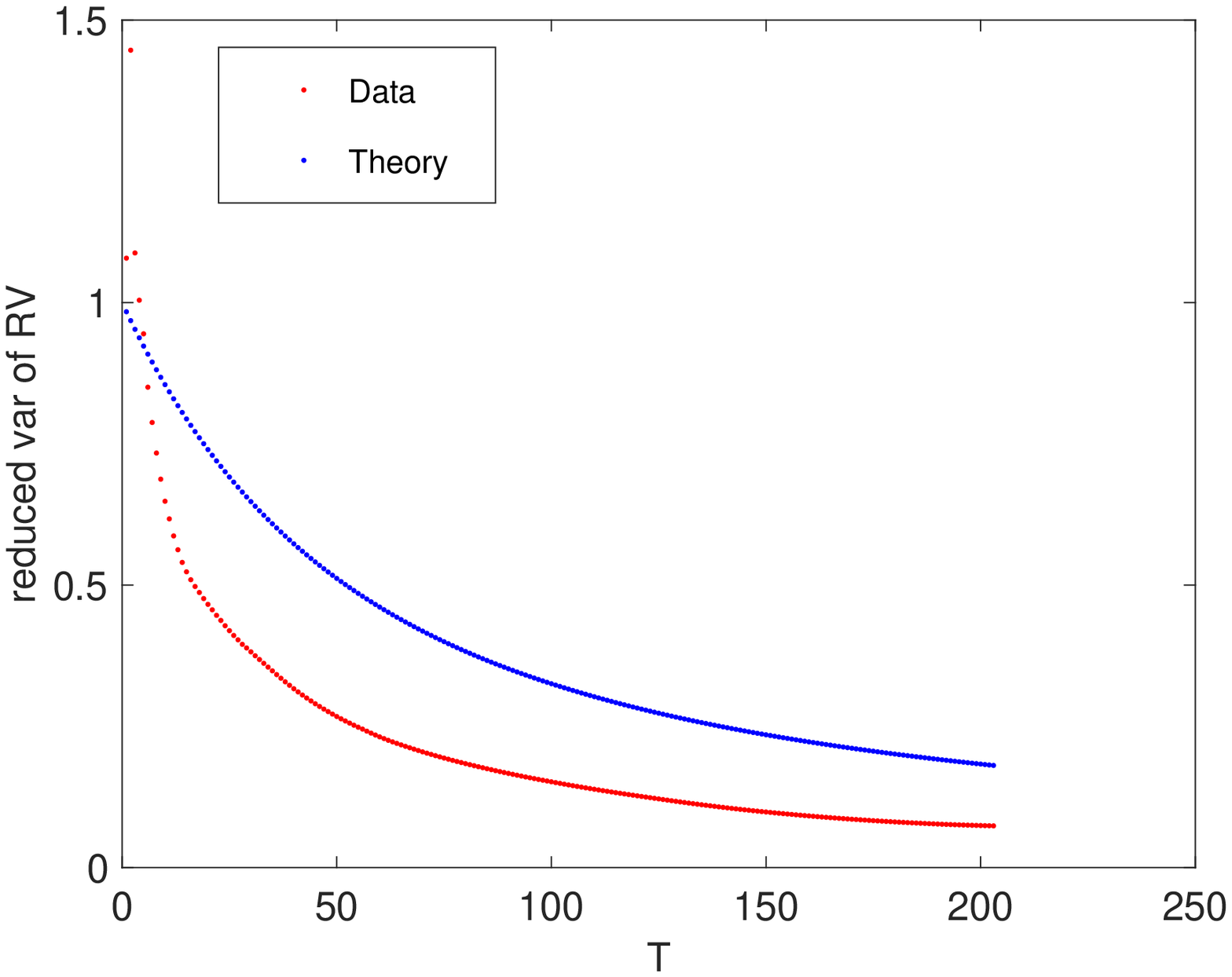}
\includegraphics[width = 0.5 \textwidth]{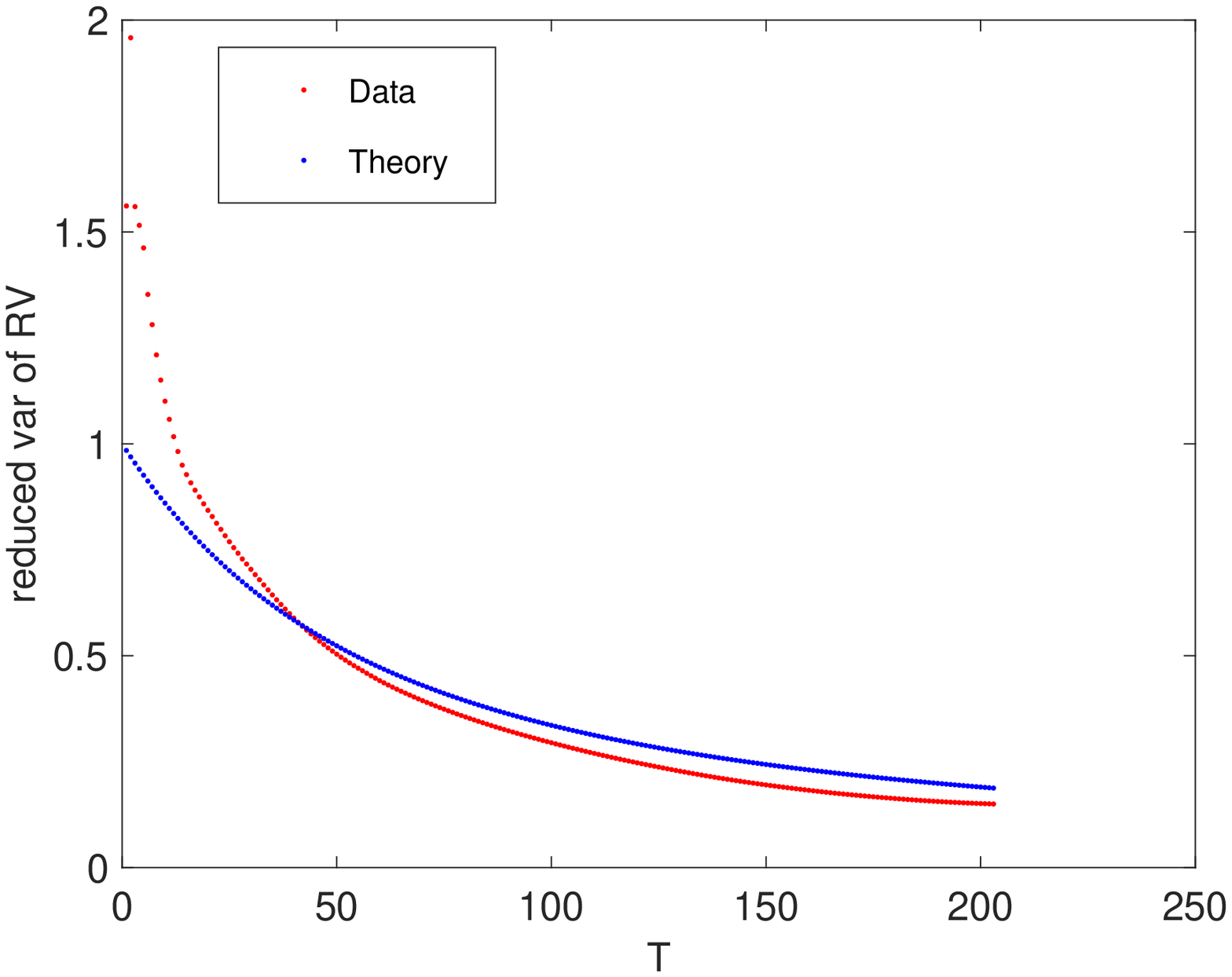}
\end{tabular}
\caption{Function $f(\gamma \tau)$ (\ref{limits}) vis-a-vis market data for $E[(\frac{1}{T}\int_{0}^{T} v_t\mathrm{d}t-\theta)^2] / var[v_t]$. Straight line fits, corresponding to limits of (\ref{limits}), are shown on log-log scale: slopes are, respectively, -0.0109 and -0.991 for DJIA and  -0.0111 and -0.988 for S\&P.}
\label{fgammat}
\end{figure}

Using (\ref{meanVarCor}), we obtain the following expression for the theoretical variance of RV:
\begin{equation}
E[(\frac{1}{T}\int_{0}^{T} v_t\mathrm{d}t-\theta)^2] = var[v_t] f(\gamma T)
\label{VarRV}
\end{equation}
where $f(\gamma T)$ describes the time dependence of the variance of RV:
\begin{equation}
f(\gamma T) = \frac{2(-1+e^{-\gamma T} +\gamma T)}{(\gamma T)^2} \approx
\begin{cases}
1  \hspace{0cm} \text{,} \hspace{.1cm} \gamma T \ll 1\\
2 (\gamma T)^{-1} \hspace{0cm} \text{,} \hspace{.1cm} \gamma T \gg 1
\end{cases}
\label{limits}
\end{equation}
We evaluate the variance of RV and $var[v_t]$ from the market data and plot their ratio, together with $f(\gamma T)$, in Fig. \ref{fgammat}. We should mention that theoretical plots with $\gamma$ found from correlations (see values in Fig. \ref{kappas}) and leverage, $\gamma_{\text{DJIA}} = 0.049, \gamma_{\text{S\&P}} = 0.047$, are virtually indistinguishable to an eye; still leverage data give about 0.6\% better fit to the market data.

\section{Conclusions \label{Conclusions}}
Multiplicative and Heston model are simple stochastic volatility models, which successfully explain many features of stock returns, particularly over multiple days of accumulation. However they suffer from some shortcomings: multiplicative model seems to underestimate the effects of volatility for small volatilities and Heston the effects for large volatilities. The combined multiplicative-Heston model studied here breeches the two models and reproduces the power-law tails of the multiplicative model for large volatilities and Heston model behavior at small volatilities.

We also examined the even moments of the stock returns vis-a-vis the theoretical predictions of this model and found a good agreement. We discussed the fact that the theoretical moments can be derived alternatively from the stock returns distribution function and stochastic variance distribution function. Towards this end, the distribution function of stock returns is best described by the product distribution of stochastic volatility distribution function and normal distribution, indicating that the stock returns equations should be interpreted in the Stratonovich sense.

Finally, we examined the correlation function of stochastic variance and used it to determine the relaxation parameter and to calculate the time dependence of the variance of realized variance. We will address the distribution of realized variance, as well as various measures of comparing it to implied variance, in a future publication \cite{dashti2018realized}.

\clearpage
\bibliography{mybib}

\end{document}